\documentclass[prb,twocolumn,showpacs,amsmath,amssymb]{revtex4}

\usepackage[pdftex]{graphicx}
\usepackage{enumerate}
\usepackage{longtable}

\begin{document}
\title{Probing orbital fluctuations in \emph{R}VO$_3$ (\emph{R}\,= Y, Gd, or Ce) by ellipsometry}
\author{J.~Reul$^{1}$}
\author{A.A.~Nugroho$^{2,3}$}
\author{T.T.M.~Palstra$^3$}
\author{M.~Gr\"{u}ninger$^{1}$}
\affiliation{$^1$ II. Physikalisches Institut,
Universit\"{a}t zu K\"{o}ln, Z\"{u}lpicher Strasse 77, D-50937 K\"{o}ln, Germany\\
$^2$ Faculty of Mathematics and Natural Sciences, Institut Teknologi Bandung, Bandung 40132, Indonesia\\
$^3$ Zernike Institute for Advanced Materials, University of Groningen, Nijenborgh 4,
9747 AG Groningen, The Netherlands}
\date{submitted: May 22, 2012; revised version: August 27, 2012}

\begin{abstract}
We study optical excitations across the Mott gap in the multiorbital Mott-Hubbard insulators \emph{R}VO$_3$.
The multipeak structure observed in the optical conductivity can be described consistently
in terms of the different 3$d^3$ multiplets or upper Hubbard bands.
The spectral weight is very sensitive to nearest-neighbor spin-spin and orbital-orbital correlations
and thus shows a pronounced dependence on both temperature and polarization.
Comparison with theoretical predictions based on either rigid orbital order or
strong orbital fluctuations clearly rules out
that orbital fluctuations are strong in $R$VO$_3$.
Both the line shape and the temperature dependence give clear evidence for the importance of
excitonic effects.
\end{abstract}

\pacs{71.27.+a, 78.40.-q, 75.25.Dk, 75.50.Mm}
\maketitle

\section{Introduction}

Orbitals play a decisive role in the low-energy physics of a large variety of transition-metal
oxides with strong electronic correlations.\cite{tokuranagaosa,dagotto}
The orbital occupation is crucial for, e.g.,\ the metal-insulator transition in V$_2$O$_3$ (Ref. \onlinecite{park2000})
and governs both the size and the sign of the exchange coupling between spins,
paving the way for, e.g.,\ low-dimensional quantum magnetism. In Mott-Hubbard insulators,
orbital degeneracy gives rise to more exotic phases such as orbital liquids.\cite{khaliullinrev}
However, the orbital degeneracy typically is lifted by the crystal field,
opening a gap for orbital excitations of a few hundred milli-electron volts or larger, strongly suppressing orbital
fluctuations. The central task in the field of orbital physics still is to establish
compounds in which the crystal field is \emph{not} dominant.
This can be achieved by, e.g.,\ strong spin-orbit coupling such as in the 5$d$ iridates,\cite{jackeli}
in which anisotropic exchange interactions may yield a realization of the Heisenberg-Kitaev model
with exotic phases and excitations.\cite{singh}
In 3$d$ systems, superexchange interactions may dominate over the crystal-field splitting.
Different groups have pointed out that orbital fluctuations may be strong in the
orbitally ordered Mott-Hubbard insulators $R$VO$_3$ ($R$\,=\,Y, rare earth),
claiming for YVO$_3$ the observation of a one-dimensional orbital liquid, of an orbital Peierls phase, and of
two-orbiton excitations.\cite{khaliullin2001,ulrich2003,horsch2003,khaliullin2004a,oles2005,oles2007,blake2009,benckiser}
Studies based on LDA+$U$ and LDA+DMFT\cite{fang2004,deRay2007} (local density approximation + dynamical mean-field theory) find that orbital fluctuations are
suppressed in YVO$_3$ by a sizable crystal-field splitting but support strong orbital fluctuations
for larger $R$ ions such as in LaVO$_3$.

The experimental determination of orbital fluctuations is a difficult task.
Our approach involves the analysis of the optical conductivity $\sigma(\omega)$.
Optical excitations across the Mott gap invoke microscopic hopping processes between adjacent sites.
Thus, the spectral weights (SWs) of these excitations depend sensitively on nearest-neighbor spin-spin
and orbital-orbital correlations and may show pronounced dependence on both polarization and
temperature $T$.\cite{khaliullin2004a,oles2005,kovaleva2004a,goessling2008,lee2005a,miyasaka2002,tsvetkov2004a,fang2003}
For the spin-spin correlations, this has been demonstrated in the 3$d^4$ manganites LaMnO$_3$ and
LaSrMnO$_4$.\cite{kovaleva2004a,goessling2008} There, the orbital occupation is independent of $T$ due to the large
crystal-field splitting of the $e_g$ orbitals,\cite{goessling2008}
and the spin-spin correlations govern the spectral weight.
For $R$VO$_3$ it has been predicted that orbital fluctuations have a strong impact on the $T$ dependence
of $\sigma(\omega)$,\cite{khaliullin2004a} but the experimental data are still controversial.
In LaVO$_3$, the spectral weight of the lowest peak in $\sigma_1(\omega)$
at 1.9 eV for polarization $E\!\parallel\!c\,$ shows a pronounced $T$ dependence, which has been taken
as strong evidence for orbital fluctuations. \cite{miyasaka2002,khaliullin2004a}
However, the proposed multiplet assignment of the peaks fails to describe the $T$ dependence of
the higher-lying peaks.
In addition, there is a striking disagreement concerning the data
reported for the sister compound YVO$_3$ (Refs. \onlinecite{miyasaka2002,tsvetkov2004a}, and \onlinecite{fujioka})
as none of the different data sets is in agreement with the theoretical
predictions.\cite{khaliullin2004a,oles2005,fang2003}

Here, we report on a detailed analysis of the optical conductivity $\sigma(\omega)$ of
YVO$_3$, GdVO$_3$, and CeVO$_3$ in the frequency range from 0.8 to 5.0 eV
for temperatures from 20 K to 500 K.\@
Our results clarify the striking discrepancies of the data reported in
Refs.\ \onlinecite{miyasaka2002,fujioka,tsvetkov2004a}
for YVO$_3$. We derive a consistent description of the observed absorption bands in terms of
the different upper Hubbard bands of these multiorbital compounds.
The temperature and polarization dependences of the spectral weights are in excellent agreement with predictions
for rigid orbital order.
We firmly conclude that orbital fluctuations are only weak in $R$VO$_3$
and propose that the lowest peak is caused by excitonic effects.

\section{Experiment}

Single crystals of \emph{R}VO$_3$ with \emph{R}\,=\,Y, Gd, and Ce
were grown by the traveling-solvent floating-zone method.\cite{blake2002}
The purities, stoichiometries, and single-phase
structures of the crystals were checked by x-ray diffraction and
thermogravimetry.
Ellipsometric data were obtained using a rotating-analyzer ellipsometer (Woollam VASE)
equipped with a retarder between polarizer and sample.
The complex optical conductivity $\sigma^j_1(\omega) + i \sigma^j_2(\omega)$ for $j \in \{a,c\}$
was derived from a series of measurements with different orientations of a polished $ac$ surface.

\section{Orbital order}

At room temperature, $R$VO$_3$ exhibits an orthorhombic crystal structure
(\textsl{Pbnm}).\cite{blake2002,reehuis2006,ren2003,sage2007}
A phase transition to a monoclinic phase (\textsl{P2$_1$/b}) is
observed at $T_{OO}$\,$\approx$\,200 K ($R$ = Y), 208 K (Gd),
and 154 K (Ce),
and antiferromagnetic order sets in at $T_{N}$\,$\approx$\,116 K (Y), 122 K (Gd),
and 134 K (Ce).\cite{blake2002,ren2000,kawano1994,ren2003,miyasaka2003}
YVO$_3$ shows a further phase transition at $T_S$\,=\,77 K to a low-temperature orthorhombic
phase,\cite{blake2002,reehuis2006} which is absent for $R$ = Gd and Ce.
We use the same set of axes at all temperatures
and neglect the small monoclinic distortion.\cite{blake2002}
The undoped Mott-Hubbard insulators \emph{R}VO$_3$ have two electrons in the $3d$ shell per V site.
In the ground state, both electrons occupy $t_{2g}$ orbitals with total spin 1.
The $t_{2g}$ manifold is split into $d_{xy}$, $d_{xz}$, and $d_{yz}$ orbitals
by the crystal field, and
the total splitting is on the order of 0.1 -- 0.2 eV.\cite{benckiser,deRay2007,solovyev2008,otsuka2006}
In the orbitally ordered phases, the $d_{xy}$ orbital is occupied by one electron
at each V site. The occupation of $d_{xz}$ and $d_{yz}$ by the second electron can be viewed as a pseudo-spin,
and both spins and pseudo-spins have been reported to show ordering patterns of either the \emph{G} type
(antiferro along all bonds, i.e.\ $xz$ and $yz$ alternate) or the \emph{C} type (ferro along $c$, antiferro
within the $ab$ plane) [see Fig.\ 1(e) and 1(f)].
In YVO$_3$, one finds \emph{G}-type spin order (SO) and \emph{C}-type orbital order (OO) below $T_S$\,=\,77 K
[see Fig.\ 1(f)].\cite{blake2002,reehuis2006,ren2000,miyasaka2003,tsvetkov2004a,noguchi2000,deRay2007}
In the monoclinic phase, all compounds show \emph{C}-type SO below $T_N$, while the correct description
of the orbitals is controversial (see the discussion in Ref.\ \onlinecite{benckiser}).
Both, \emph{G}-type orbital order \cite{noguchi2000,blake2002} [see Fig.\ 1(e)] and strong orbital
fluctuations\cite{khaliullin2001,ulrich2003,khaliullin2004a,oles2005,horsch2003,oles2007,blake2009}
have been claimed. Here, we show that the optical data rule out strong orbital fluctuations.

\begin{figure}[tb]
\includegraphics[width=1.0\columnwidth,clip]{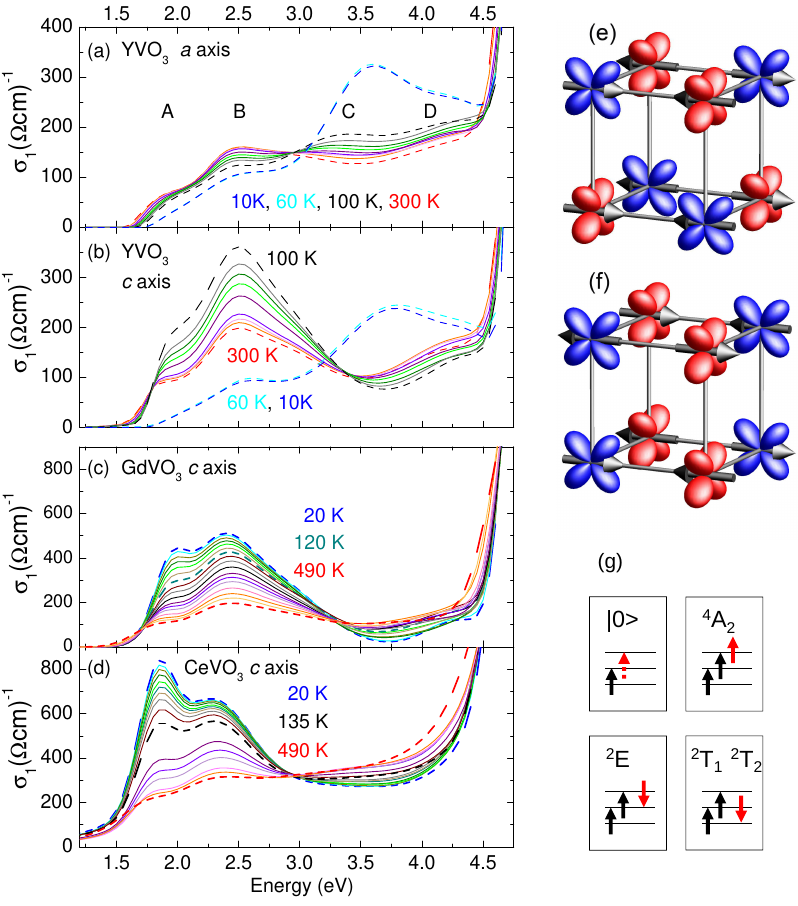}
\caption{(Color online) (a) Optical conductivity $\sigma_1^a(\omega)$ of YVO$_3$.
(b)-(d) $\sigma_1^c(\omega)$ of YVO$_3$, GdVO$_3$, and CeVO$_3$, respectively. Temperature labels refer
to the dashed lines. The absorption bands are labeled as peaks A ($\approx\,2$ eV), B ($\approx\,2.5$ eV),
C (3.0 - 3.6 eV), and D (3.8 - 4.4 eV).
(e) \emph{G}-type OO with \emph{C}-type SO.
(f) \emph{C}-type OO with \emph{G}-type SO.
(g) Sketch of $t_{2g}^2$ and $t_{2g}^3$ multiplets.
}
\label{fig:YGdCeVO3}
\end{figure}

\section{Results}
\subsection{Optical conductivity and multiplet assignment}

The overall behavior of $\sigma_1(\omega)$ is very similar for YVO$_3$, GdVO$_3$, and CeVO$_3$
[see Fig.\ 1(a) - 1(d)]. The main difference is that YVO$_3$ shows the low-temperature phase with \emph{C}-type OO
below $T_S$ = 77 K with a pronounced peak at 3.5 eV.\@
In all three compounds, the Mott gap is about 1.6 to 1.8 eV in excellent agreement with infrared-transmittance data.\cite{benckiser}
Above the gap we observe Mott-Hubbard (MH) excitations, i.e., excitations from a $|d_i^2d_j^2\rangle$ ground state
to a $|d_i^1 d_j^3\rangle$ final state, where $i$ and $j$ denote different V sites and
$d^1$ and $d^3$ refer to the lower and upper Hubbard bands, respectively.
Above 4.5 eV, $\sigma_1(\omega)$ steeply rises up to roughly $1500\,(\Omega {\rm cm})^{-1}$ (not shown),
corresponding to the onset of charge-transfer excitations from the O$_{2p}$ band to the upper Hubbard band,
$|d^2p^6\rangle \rightarrow |d^3p^5 \rangle$.
This general picture is well accepted.\cite{miyasaka2002,fujioka,tsvetkov2004a,fang2003,solovyev1996}
The MH excitations show a multipeak structure (peaks A - D at roughly 2 eV, 2.5 eV, 3 - 3.6 eV,
and 3.8 - 4.4 eV, respectively), which is expected to reflect the local $d^3$ multiplet structure of these multiorbital
systems.\cite{miyasaka2002,tsvetkov2004a,oles2005,khaliullin2004a,fang2003}
The splitting between the $t_{2g}$ level and the $e_g$ level amounts to 10\,Dq$\,=\,1.9$ eV in YVO$_3$.\cite{benckiserDiss}
For a discussion of the lowest excited states we thus may neglect the $e_g$ orbitals.
For the sake of simplicity, we assume cubic symmetry and neglect the crystal-field splitting within the $t_{2g}$ levels
of roughly 0.1 - 0.2 eV.\cite{benckiser,deRay2007,solovyev2008,otsuka2006}
In the ground state, the $t_{2g}^2$ configuration shows $^3T_1$ symmetry with spin 1.
The excited states $|t_{2g}^1t_{2g}^3\rangle$ have to be distinguished according to the $t_{2g}^3$ sector,
because the $t_{2g}^1$ configuration always has the same energy.
The $t_{2g}^3$ multiplets exhibit energies of
$U - 3 J_H$ ($^4A_2$), $U$ ($^2E$, $^2T_1$), and $U + 2J_H$ ($^2T_2$)
[see Fig.\ 1(g)]\cite{oles2005,tsvetkov2004a,khaliullin2004a}
with the on-site Coulomb repulsion $U$\,$\approx \, $4 -- 5 eV (Refs. \onlinecite{deRay2007} and \onlinecite{mizokawa1996})
and the Hund coupling $J_H \approx 0.55 - 0.7$ eV.\cite{benckiser}

We now focus on a consistent assignment of the MH excitations to the different $t_{2g}^3$ multiplets.
The spectral weight of a given excitation depends on the spin-spin and orbital-orbital correlations
between adjacent sites and thus depends strongly on polarization and temperature.
The lowest multiplet $^4A_2$ is a high-spin state in which the $xy$, $xz$, and $yz$ orbitals are
occupied by one electron each [see Fig.\ 1(g)]. Due to the high-spin character, parallel spins on adjacent sites in the
initial state give rise to a larger SW than antiparallel spins. In contrast, the other $t_{2g}^3$ multiplets
$^2E$, $^2T_1$, and $^2T_2$ all are low-spin states, thus the SW is larger for \emph{antiparallel} spins.
This yields the following clear predictions for the phase with \emph{C}-type SO\cite{khaliullin2004a,oles2005,fang2003}
in which spins are parallel along the $c$ axis and antiparallel within the $ab$ plane [see Fig.\ 1(e)].
(1) The SW of the excitation into the lowest multiplet $^4A_2$ is expected to be larger
in $\sigma^c_1$ than in $\sigma^a_1$.
(2) With decreasing temperature $T$, spin-spin and orbital-orbital correlations are enhanced,
thus $\sigma^c_1$ ($\sigma^a_1$) is expected to increase (decrease) for the lowest multiplet.
(3) The \emph{opposite} $T$ dependence is expected for the higher multiplets.
A comparison of these predictions with our data clearly shows that \emph{both} peaks A and B at 2.0 and 2.5 eV, respectively,
have to be assigned to the lowest $3d^3$ multiplet $^4A_2$. Peak C is located roughly $3J_H$ above peak A
in $\sigma^a_1$, in agreement with the expectations for the ($^2E$, $^2T_1$) multiplets.
The SW of this excitation vanishes for parallel spins,\cite{khaliullin2004a,oles2005,fang2003}
and therefore, it is absent in $\sigma^c_1$ in the phase with \emph{C}-type SO.

The dramatic changes observed at $T_S$ = 77 K in YVO$_3$ unambiguously prove that our peak assignment
is correct.
At $T_S$, the nearest-neighbor correlations along $c$ change from ferro to antiferro for the spins
and vice versa for the orbitals, thus two adjacent sites show the same orbital occupation below $T_S$
with, e.g.,\ $xz$ occupied on both sites. In this case, an excitation to $^4A_2$ requires hopping
from $xz$ on one site to $yz$ on a neighboring site, which is forbidden along $c$ in cubic symmetry,
explaining the spectacular suppression of peaks A and B. The finite SW at low $T$ is due
to deviations from cubic symmetry.\cite{fang2003}
At the same time, the transition to the $^2T_1$ multiplet (contributing to peak C) strongly favors
\emph{G}-type SO and \emph{C}-type OO, explaining the dramatic increase of peak C below 77 K.\@

The highest $t_{2g}^3$ multiplet $^2T_2$ is roughly expected at $U + 2J_H$, i.e.,\ $5 J_H\,   > \, 2.7$ eV
above the lowest peak. It is thus reasonable to assume that this excitation is located above the onset
of charge-transfer excitations at 4.5 eV.\@
Peak D at 3.8 - 4.4 eV lies about 10\,Dq = 1.9 eV (Ref. \onlinecite{benckiserDiss}) above peaks A and B and
thus can be assigned to the lowest $t_{2g}^2 e_g^1$ multiplet. In cubic symmetry, the excitation from
a $t_{2g}$ orbital on site $i$ to an $e_g$ orbital on a neighboring site is forbidden,
but deviations from cubic symmetry yield a finite spectral weight. Accordingly, peak D is hardly visible
in less-distorted CeVO$_3$.

\subsection{Comparison with literature}

Conflicting with our assignment, peaks A and B have been attributed to the two lowest multiplets
$^4A_2$ and ($^2E$, $^2T_1$) in Refs.\ \onlinecite{miyasaka2002,tsvetkov2004a,fujioka,fang2003}.
Between these multiplets a splitting of $3 J_H > 1.5$ eV is expected, which is incompatible
with the observed splitting between A and B of only 0.5 eV.\@
In other words, the previous assignment of peaks A and B to two different multiplets
yields a nonphysically small value of $J_H$.\cite{fang2003}
Moreover, this scenario is inconsistent
with the fact that the SWs of peaks A and B show the same $T$ dependence, as discussed above.
Additionally, the data for YVO$_3$ are strikingly different.\cite{miyasaka2002,tsvetkov2004a}
In Ref.\ \onlinecite{tsvetkov2004a}, the pronounced peak at 3.5 eV, characteristic of the low-temperature
phase, is also observed above $T_S$, whereas it is not seen at any temperature in Ref.\ \onlinecite{miyasaka2002}.
Both the incorrect assignment and the discrepancies of the data can be traced back to problems
with the sample temperature. Samples of YVO$_3$ tend to break at the first order structural transition
at $T_S$, often leading to a loss of thermal contact.\cite{fujioka} We were able to avoid this problem
by very slow cooling. A comparison of our data and the data of Ref.\ \onlinecite{fujioka} shows that
the seemingly contradictory data of Refs.\ \onlinecite{miyasaka2002} and \onlinecite{tsvetkov2004a} can be reconciled with
each other by taking into account problems with the sample temperature across $T_S$.
The data of Ref.\ \onlinecite{fujioka} show the expected jump of the spectral weight at $T_S$,
but both this jump and the $T$ dependence above $T_S$ are much smaller than in our data.
We attribute this difference to the different experimental techniques. Reference \onlinecite{fujioka}
reports reflectivity data with a subsequent Kramers-Kronig analysis. In contrast, ellipsometry
is a self-normalizing technique which is much better suited for a precise determination of
the $T$ dependence.\cite{kovaleva2004a,tobe2001}

\begin{figure}[tb]
\includegraphics[width=0.8\columnwidth,clip]{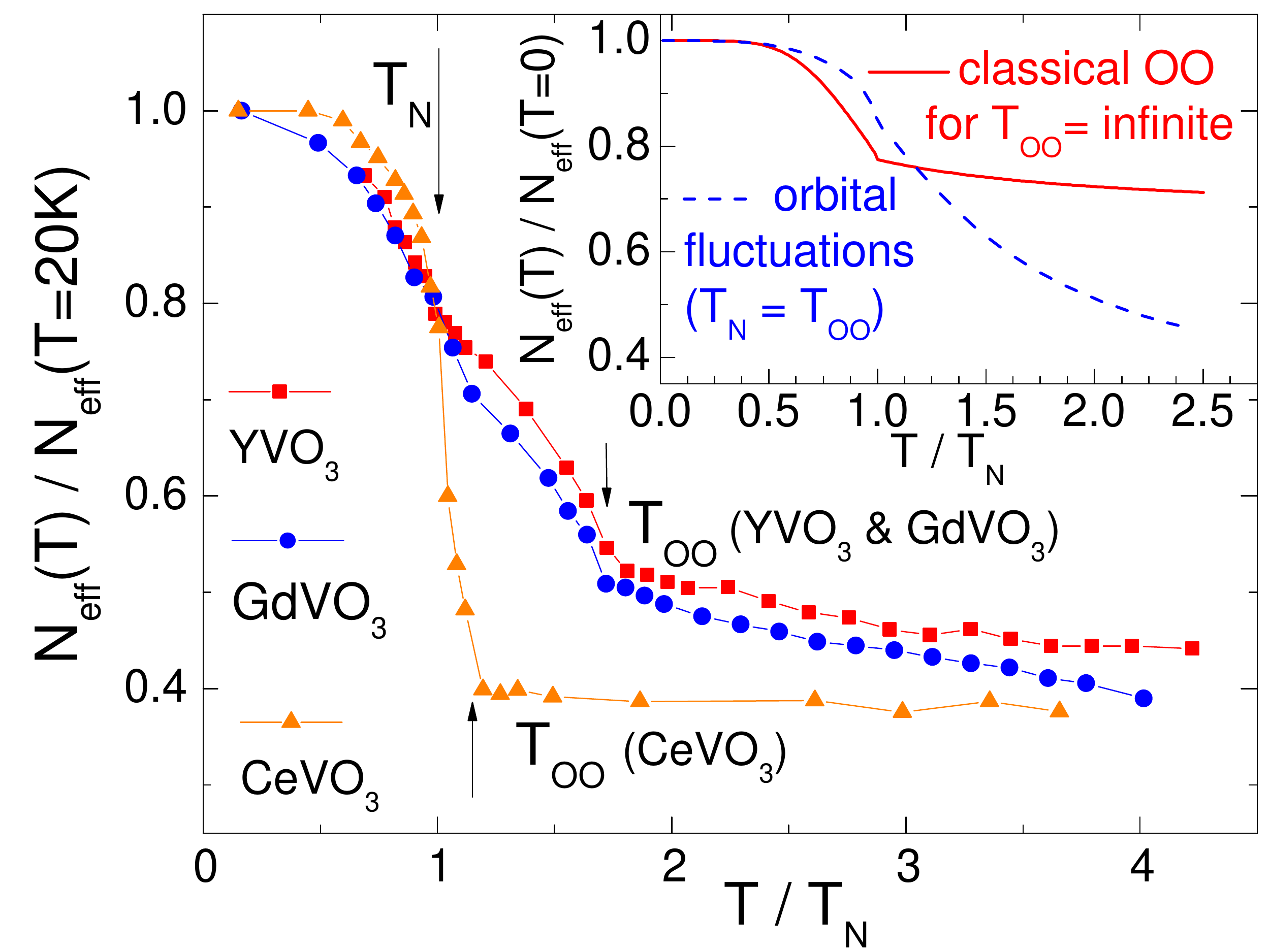}
\caption{(Color online) Sum of the spectral weights of peaks A and B in $\sigma_1^c$.
Inset: Theoretical results for $N_{\rm eff}(T)$ of the lowest Mott-Hubbard excitation ($^4A_2$ multiplet) in $\sigma_1^c$
for strong orbital fluctuations (blue, $T_N = T_{OO}$) and rigid orbital order (red, $T_{OO} = \infty$, i.e.,\ only the
reduction of the spin part to 2/3 is taken into account [see Refs. \onlinecite{khaliullin2004a} and \onlinecite{oles2005}]).
}
\label{fig:theory}
\end{figure}

\subsection{Temperature dependence of the spectral weight: strength of orbital fluctuations}

The correct assignment of peaks A and B to the lowest multiplet $^4A_2$ is crucial for the discussion
of the role of orbital fluctuations.
The spectral weight of the $^4A_2$ multiplet in $\sigma_1^c$ depends sensitively on spin-spin and orbital-orbital
correlations between adjacent sites.\cite{khaliullin2004a,oles2005,fang2003}
Comparing a fully polarized ferromagnetic state ($T$=0) with a disordered spin state ($T$=$\infty$),
the SW in the latter case is reduced to 2/3 based on the nearest-neighbor hopping matrix element.
The change from a fully ordered orbital state to a disordered one yields another factor 1/2.
Thus, in total one expects a reduction of the SW by a factor of 3 from low $T$ to high $T$.\cite{khaliullin2004a,oles2005}
This is valid in any scenario,
i.e., it applies to both rigid orbital order and strong orbital fluctuations.
In order to determine the strength of orbital fluctuations, one has to study
the detailed $T$ dependence of the SW, which thus is the most interesting quantity.
The inset of Fig.\ 2 shows the predictions for both strong orbital fluctuations (blue line,
assuming $T_{OO}=T_N$) and rigid orbital order (red line, assuming $T_{OO}=\infty$).
A near coincidence of the ordering temperatures $T_{OO}$ and $T_N$ for orbitals and spins is realized in CeVO$_3$.
The red line for rigid orbital order shows only the reduction by a factor of two stemming from the spin part,
because it assumes $T_{\rm OO} = \infty$. The key feature of this comparison is not the difference in absolute
value but the $T$ dependence above the ordering temperature.
For rigid orbital order, the SW is nearly constant for $T>T_N$ and exhibits a clear kink right at $T_N$.
In contrast, there is no pronounced effect at $T_N$ for strong quantum fluctuations.
The smoking gun for strong fluctuations is a strong $T$ dependence far above $T_N$ or $T_{OO}$;
most of the change occurs above the ordering temperature.

In Fig.\ 2, these predictions are compared with our results. The SW of a
single absorption band is given by $N_{eff}=(2mV/\pi e^2)\int^\infty_0\sigma_1(\omega)d\omega$,
where $m$ is the free-electron mass and $V$ is the volume per magnetic ion.
We used four Lorentz oscillators to describe peaks A - D.
The total SW of peaks A and B is shown in Fig.\ 2.
For all compounds we find nearly constant spectral weight above T$_{OO}$, a clear kink at $T_{OO}$,
and also a kink at T$_N$. These findings are in excellent agreement with the expectations for
rigid orbital order.
The fact that the changes above $T_{OO}$ are much smaller than below rules out strong orbital fluctuations.
Also, the total change of the SW is in excellent agreement with theory.

The claim that orbital quantum fluctuations are strong in pseudocubic $R$VO$_3$ is based on the idea that
superexchange interactions between $t_{2g}$ orbitals are frustrated on a cubic lattice.\cite{khaliullin2001}
More precisely, orbital quantum fluctuations are blocked in the $ab$ plane due to the occupied $d_{xy}$ orbital,
but they have been claimed to be strong along the $c$ axis in the monoclinic phase where orbital fluctuations and
ferromagnetic spin order may support each other.\cite{khaliullin2001}
However, distortions away from cubic symmetry give rise to both a crystal-field splitting of the $t_{2g}$ orbitals
and a reduction of superexchange interactions. The orthorhombic splitting $\varepsilon \,= \,(b\,-\,a)/(b\,+\,a)$
between the lattice parameters $a$ and $b$
amounts to 0.03 for $R$=Y and Gd and only 0.003 for Ce \cite{reehuis2006,sage2007,munoz2003}
while the V-O-V bond angle increases from about 144$^\circ$ in YVO$_3$ to about 156$^\circ$ in
LaVO$_3$.\cite{blake2002,sage2007,bordet1993}
Our results clearly show that orbital fluctuations
are suppressed not only in strongly distorted YVO$_3$ but also for large $R$ ions such as in pseudocubic CeVO$_3$.

\begin{figure}[tb]
\includegraphics[width=0.66\columnwidth,clip]{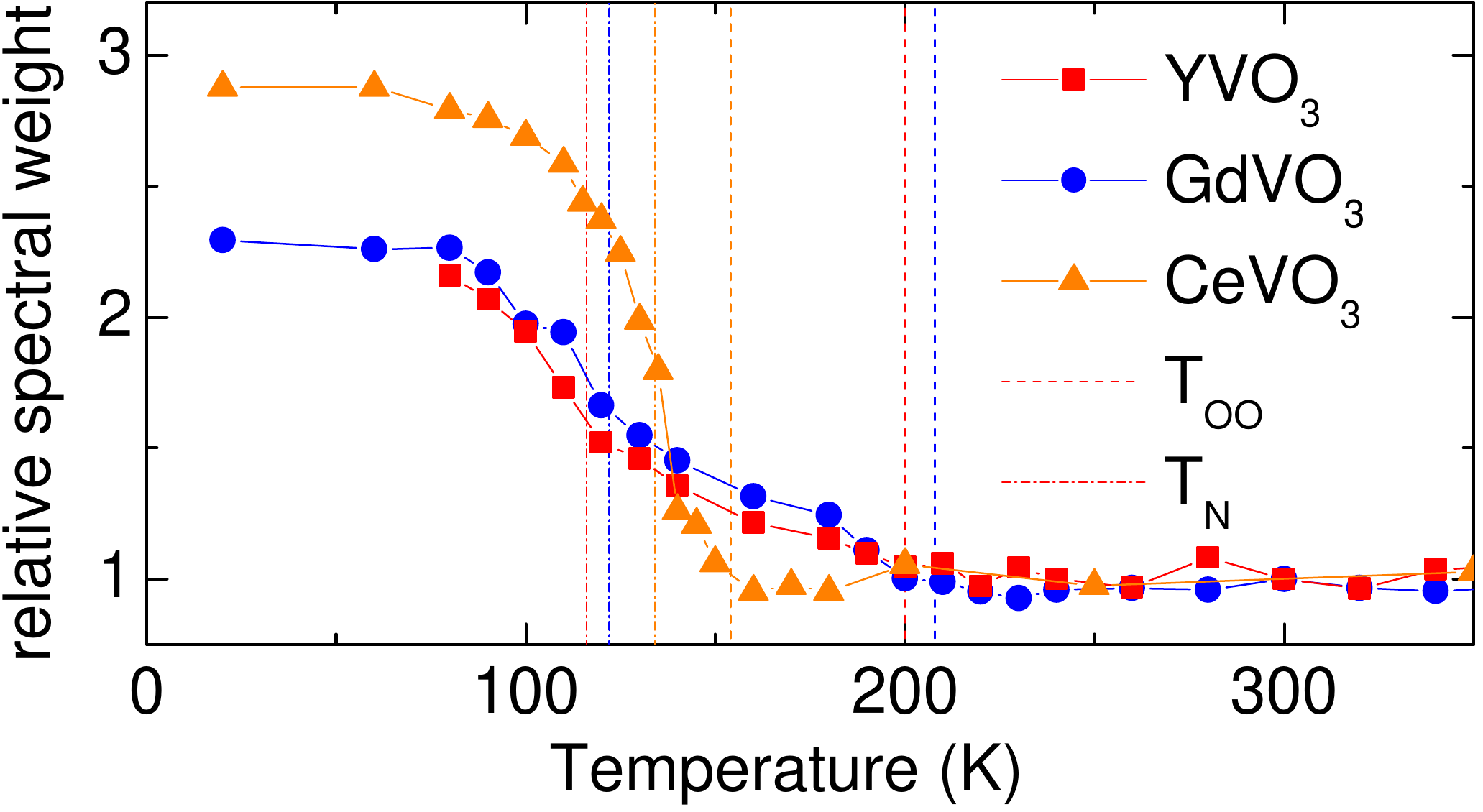}
\caption{(Color online) Ratio of the spectral weights of peaks A and B in $\sigma_1^c$,
normalized to the value at 300 K. }
\label{fig:3}
\end{figure}

\subsection{Hubbard exciton}

Finally, we address the double-peak structure A/B of the excitation into the lowest multiplet.
Similar double- and multi-peak structures of the lowest multiplet have been reported for YTiO$_3$
and LaMnO$_3$.\cite{goesslingTi,kovaleva2004a} The peak splitting has been assigned to either
excitonic or band-structure effects, which both have been neglected thus far
in our discussion of local multiplets.
We propose that peak A is an excitonic resonance, i.e., not a truly bound state below
the gap but a resonance within the absorption band. Due to an attractive interaction between
a $d^3$ state in the upper Hubbard band and a $d^1$ state in the lower Hubbard band, the energy of
the resonance (peak A) is less than the energy of peak B where peak B reflects an excitation
to $|d^1 d^3\rangle$ without attractive interaction. In order to substantiate this
claim, we discuss results from photoemission spectroscopy (PES)\cite{maiti2000,mossanek2008}
and from band-structure calculations.\cite{fang2003,solovyev1996,deRay2007}

Using LDA+$U$, Fang \emph{et al.}\cite{fang2003} calculated the optical conductivity of LaVO$_3$ and
YVO$_3$ for different polarizations and for different ordering patterns of spins and orbitals.
The two lowest peaks (called $\alpha$ and $\beta$ in Ref.\ \onlinecite{fang2003}) correspond
to the two lowest multiplets of our local approach, i.e.,\ to the double-peak A/B and C, respectively.\@
Accordingly, the spectral weight of peak $\alpha$ ($\beta$) in $\sigma_1^c$ decreases (increases) in YVO$_3$
across the phase transition from the intermediate phase with \emph{C}-type SO to the low-$T$
phase with \emph{G}-type SO, as observed experimentally for peak A/B (C).\@
Peak $\alpha$ is the lowest peak, well separated from the higher-lying excitations, and clearly
consists of a single peak only, both for YVO$_3$ and LaVO$_3$.\cite{fang2003} The experimentally
observed splitting into peaks A and B is absent in the LDA+$U$ results, which neglect
excitonic effects. For the intermediate phase of YVO$_3$, Fang \emph{et al.}\cite{fang2003} predict
peaks $\alpha$ and $\beta$ (with $\beta$ observable for $E\perp c$ only) at about 1.7 and 2.9 eV, respectively,
both significantly lower than peaks B and C in experiment, but in LDA+$U$ results, the peak energies depend
sensitively on the particular choice of $U$. Considering only the energies, one may be tempted to
assign peaks $\alpha$ and $\beta$ to peaks A and B, but this is clearly ruled out by the dependence
of the spectral weight on both temperature and polarization as well as by the value of $J_H$,
as discussed in Sec. IV A.\@
Also the LDA+$U$ study of Solovyev \emph{et al.}\cite{solovyev1996} reports on the optical
conductivity of LaVO$_3$. In agreement with the results of Fang \emph{et al.},\cite{fang2003}
there is no splitting of the lowest excitation.
Solovyev \emph{et al.}\cite{solovyev1996} find the band gap at 0.7 eV and the charge-transfer gap
at about 3.5 eV while the lowest absorption band is peaking at about 1.7 eV.\@ Since both gaps are
about 0.7 -- 1 eV lower than in experiment, we assign the lowest peak from LDA+$U$ at 1.7\,eV
to peak B in our data. As mentioned above, the peak energy depends sensitively on the choice of $U$.
The LDA+DMFT study of De Raychaudhury \emph{et al.}\cite{deRay2007} does not report on the optical
conductivity, but it shows the electron-removal and -addition spectra (as measured by PES and
inverse PES) for LaVO$_3$ and YVO$_3$. For LaVO$_3$, the electron-removal spectrum shows contributions
from all three $t_{2g}$ orbitals, peaking at about 1.2 -- 1.4 eV below the Fermi energy $E_F$.\cite{deRay2007}
The small splitting reflects the crystal-field splitting within the $t_{2g}$ level.
The first peak of the electron-addition spectrum lies at about 1.2 eV above $E_F$. Neglecting excitonic
effects, one may thus expect the first peak in the optical conductivity at about 2.4 -- 2.6 eV,
which is in agreement with peak B.\@ For YVO$_3$, the electron-removal and -addition spectra peak at about
-1.4 to -1.5 eV and +1.2 eV, respectively; \cite{deRay2007} thus, the peak in the optical conductivity is expected
at a slightly larger energy in YVO$_3$ than in LaVO$_3$, in agreement with our experiment.
The calculated electron-removal and -addition spectra for YVO$_3$ show small shoulders
at about -1.1 and +0.7 eV.\@ However, similar features are absent in the calculated spectra of LaVO$_3$.
In strong contrast, peak A in the optical conductivity is much more pronounced in LaVO$_3$ than in YVO$_3$.
In summary, band-structure calculations do not provide any explanation for the observed splitting
of about 0.5 eV between peaks A and B.\@

Experimental photoemission spectra of LaVO$_3$ and YVO$_3$ show a single peak
lying about 1.5 -- 1.8 eV  below $E_F$.\cite{maiti2000,mossanek2008}
For LaVO$_3$, the combination of PES and inverse PES has been reported by Maiti and Sarma.\cite{maiti2000}
The separation between the highest peak below $E_F$ and the lowest peak above $E_F$ amounts to roughly 3 eV,
but the inverse PES data were measured with a resolution of only 0.8 eV.\@
These results are in agreement with the LDA+DMFT study of Ref.\ \onlinecite{deRay2007} discussed above.
Neither band-structure calculations nor the PES data provide any explanation for the splitting
of peaks A and B.\@ Electron-removal and -addition spectra do not reflect excitonic effects
in contrast to the optical conductivity. Altogether, this strongly supports an excitonic
interpretation of peak A.\@

In simple band insulators, exciton formation is driven by a lowering of the Coulomb energy
whereas the kinetic energy increases.
The term \emph{Hubbard exciton} refers to an exciton in a Mott-Hubbard insulator.
Such Hubbard excitons are of particular interest because of the possible role of the
\emph{kinetic} energy for the attractive interaction.\cite{clarke,wang,zhangng,wrobel,kuzian}
In $R$VO$_3$, the ratio SW$_{\rm A}$/SW$_{\rm B}$ of the spectral weights of peaks A and B in $\sigma_1^c$ strongly increases
from $R$=Y via Gd to Ce. Interestingly, this ratio also depends sensitively on the temperature, but only below
the orbital-ordering temperature $T_{\rm OO}$ (see Fig.\ 3). Below $T_{\rm OO}$, the SWs of both peaks A and B increase,
but this increase is much more pronounced for the excitonic peak A.\@
This clearly demonstrates the significant role of orbital order for exciton formation
in Mott-Hubbard insulators. We propose that this influence of orbital order indicates the importance of
the kinetic energy for Hubbard excitons
in the case of antiferro-orbital order (here, along $c$), as discussed for YTiO$_3$.\cite{goesslingTi}
Along $c$ ,
the motion of an exciton is not hindered by the antiferro-orbital order
whereas the hopping of either a single $d^3$ state or a single $d^1$ state leaves a trace of orbitally
excited states (see Ref.\ \onlinecite{goesslingTi} for a more detailed discussion).
This results from the restriction that hopping is essentially only allowed within the same type of
orbital, e.g.,\ from $xz$ on one site to $xz$ on a neighboring site.
Therefore, the exciton can hop on a larger energy scale than a single $d^1$ state or a single $d^3$ state,
and exciton formation is equivalent to a {\it gain of kinetic energy}.\cite{goesslingTi}
This scenario
is supported by recent pump-probe measurements on YVO$_3$, which cover the frequency range of
peaks A and B.\cite{novelli}

\section{Summary and conclusions}

In summary, we provide a consistent assignment of the Mott-Hubbard excitations and a quantitatively reliable
$T$ dependence of the spectral weights of YVO$_3$, GdVO$_3$, and CeVO$_3$.
A comparison of our data with theoretical results\cite{khaliullin2004a,oles2005} clearly rules out
strong orbital fluctuations in $R$VO$_3$.
We propose that the line shape and the $T$ dependence of the lowest absorption band reflect excitonic effects.

\section*{ACKNOWLEDGMENTS}

We thank G. Khaliullin, P. Horsch, and E. Pavarini for fruitful discussions.
A. A. N. acknowledges support from the I-MHERE staff-exchange program, FMIPA-ITB.\@
This work was supported by the DFG via SFB 608 and BCGS.

\end{document}